УДК 004.051

# ДОВЕРЕННАЯ МАРШРУТИЗАЦИЯ ПРОТИВ VPN ЗА БЕЗОПАСНОСТЬ ПЕРЕДАЧИ ДАННЫХ ПО IP-СЕТЯМ/ИНТЕРНЕТ

*Тиамийу Абдулрахамон Осуолале*

*аспирант Санкт-Петербургского государственного университета телекоммуникаций имени проф. М.А. Бонч-Бруевича, г. Санкт-Петербург*

**АННОТАЦИЯ**

Хотя цели у доверенной маршрутизации и виртуальных частных сетей совпадают – обеспечение безопасной передачи данных от отправителя к получателям через общедоступные сети, такие как Интернет, все же между этими двумя методами имеются существенные различия. В этой статье анализируются их различия.

**Ключевые слова:** доверенная маршрутизация; VPN; защита данных; контроль; вторжение.

# TRUSTED ROUTING VS. VPN FOR SECURED DATA TRANSFER OVER IP-NETWORKS/ INTERNET

*Tiamiyu A. Osuolale*

*Postgraduate student of the Bonch-Bruevich Saint-Petersburg State University of Telecommunications, Saint Petersburg*

**ABSTRACT**

Though objectives of trusted routing and virtual private networks (VPN) data transfer methods are to guarantee data transfer securely to from senders to receivers over public networks like Internet yet there are paramount differences between the two methods. This paper analyses their differences.

**Keywords**: trusted routing; VPN; data security; control; intrusion.

**Introduction**

In every organizations security objectives are availability, integrity and confidentiality popularly known as AIC triad. How a sender transfers data to the receiver safely and securely over public networks nowadays is too challenging. Intruders' knowledge base is ever-increasing and everyone, both home users and organizations are targets. Encryption of data being transferred is not discouraging the intruders from hijacking and subsequently examine/copy/modify the data being transferred especially over public networks.

**TR and VPN in brief**

TR method is detailed in ISO 7498-2-99 regarding the requirements set out in paragraph 5.3.7.1. − "Mechanisms of routing control", and this method is a reliable one that guaranty safety and security of data transfer from sender to the receiver as it avoids the possibility of data being tampered with in any form while traversing through the networks as the concept itself refers to the process of planning the transfer of information flow on the calculated route through telecommunication network nodes, excluding the possibility of substitution, modification or inclusion of any form of information into the data stream passing through those nodes [1].

A VPN extends private network services for organization(s) across a public or shared network infrastructure such as the Internet or service provider backbone network. The shared service provider backbone network is known as the VPN backbone and is used to transport traffic for multiple VPNs, as well as possibly non-VPN traffic. VPNs based on Frame Relay and ATM virtual circuits technologies have been available for a long time, but over the past few years, IP and IP/Multiprotocol Label Switching (MPLS)-based VPNs have become more and more popular [2].

**Objectives of TR and VPN**

The set of techniques and realization method of trusted routing (TR) allow for safe transfer of information from the sender to the recipient, dynamically generate routes between sender and receiver, use secure autonomous systems (ASs) for data transmission, zombifying routing devices (RD) in the absence of a route that passes only through trusted AS, control of the traffic flows and control of the operation of RD [3] while objectives of VPN are to guarantee anytime access remotely but securely to users of the VPN resources via authentication and the use of tunneling protocols and security procedures such as encryption [4].

While TR eliminates possibility of data from passing through malicious networks or potential malicious networks so as not to allow intruders having copy of the data and or tampering with the data in any form, VPN technology allows user's network traffic to pass securely over public networks using encryption and tunneling protocols. A VPN makes communications private simply by obscuring the VPN from external users so as to maintain security.

**Topology**

VPN emulates WAN using shared or public IP facilities, such as the Internet or private IP backbones. TR also uses public networks like Internet but defines and prepares secured network routes considering the locations of the senders in relation to the receivers of the data globally over Internet. While VPN can work in either a 'fully-meshed' topology or a 'non-meshed' topology, TR works well only in a fully-meshed topology.

In a fully-meshed topology, RD are connected with many redundant interconnections between network nodes. In a true meshed topology, every RD has a connection to every other RD in the network. An advantage of such a network would be that no RD is reliant upon a single connection.

In a non-meshed topology all RD are connected to a central hub that dictates the access rules of the VPN to the other network sites.

**Interconnections and Network Traffics**

TR method identifies trusted routes among interconnected RD within Internet and prepares routes base on trustworthiness of RD. For VPN, its authority will administer and serve as the main point of contact for the VPN users. And for this the authority may outsource some functions and connectivity, set up contractual agreements with the different Service Providers involved, and coordinate configuration, performance, and fault management.

**Routing Process Specifics**

TR prepares trusted routes and "Zombifies" (i.e. make trusted by gaining full control) untrusted nodes in the routes to prevent unauthorized access [5]. And then performs routing via trusted routes obtained using routing table(s) retrieved during preparation of trusted routes. Whereas VPN uses tunneling protocols, such as IP Security (IPsec), Point-to-Point Tunneling Protocol (PPTP), or Layer 2 Tunneling Protocol (L2TP), and then packaging the tunneled packet into an IP packet. The resultant packet is then routed to the destination network using the overlying IP information [6].

### Data Traffic Policy

In TR data traffic is allowed across shared or public networks only through routing devices (RD) already identified as trusted RD but this is not the case with VPN. VPN enables a host computer to send and receive data across shared or public networks as if the host were an integral part of the VPN. Data traffic is allowed through any RD that is participating in routing process globally.

### Licensing Policy

VPN ISP providers may be forced by law to allow for traffic tapping by government agency as a national security measures. Moreover VPN ISP might have its services expanded through other VPN ISP providers that may span globally, and that the knowledge of which ISP may not be willing to make public. With TR, one can avoid certain ISP (VPN server) network that might be located within unwanted locations.

### Control of Traffic Flows

TR allows for control over the information flows on trusted routes and control of the status of trusted routes by monitoring the status of the trusted RD to ensure that:

− RD regularly sends data stream according to their routing tables (RT);

− RT of RD does not change during the data transfer session;

− Labels that are in the MIB of RD do not change during the data transfer session. For traffic flow control in VPN, a Unique VPN identifier (VPN-ID) may be included in a control or data packet, to identify the "scope" of a private IP address and the VPN to which the data belongs.

### Scalability

TR may span multiple IP Autonomous Systems (AS) or Service Providers. Though a VPN may span multiple IP AS or Service Providers as well, but only as permitted by the VPN authority.

**Reliability**

VPN-ID may be included in a MIB to configure attributes to a VPN, or to assign a physical or logical access interface to a particular VPN. For reliability of TR methods, labels are included in MIB; and injection into the MIB not just one but several labels that are bound to specific router improves reliability. Moreover these labels can be changed dynamically at regular intervals.

**Data Encryption**

Labels in MIB can be encrypted to improve the reliability of TR method. Also data encryption is allowed. More emphasis is on encryption when it comes to VPN. VPN provides an authenticated, encrypted tunnel for securely transfer of data over public networks i.e. data is being encrypted and then transfer over public networks.

**Network Type**

Though TR is physically a network and not geographically dispersed, VPN is physically not a network but geographically dispersed. TR method allows sender of data to identify secured routes and then transfer data to a receiver globally within public networks while VPN allows users to connect virtually over public networks or other private IP backbones to private networks and transfer data as if they are part of the private networks e.g. branch office connecting to the headquarters' networks as if it is part of the headquarters' networks (site-to-site VPN).

**Data Security Control**

TR stops traffic flow as soon as it detects that there is changes in the RT as compared to the ones earlier retrieved during secured routes preparation thereby excluding attacks and further damages by the intruder(s). When this happens, new secured routes are defined and prepared for the transfer to continue. In contrary VPN traffic is often invisible to intrusion detection system (IDS) monitoring, and as such if VPN is intruded, the intruder can attack the internal systems without being picked up by the IDS. And as such VPN is prone to security treat as the data tunnel is rendered across multiple routers and being part of the tunnel, these intermediate routers can examine/copy/modify the data should it malicious even if the packets are encrypted.

**Top-most Priority**

Security via encryption and tunneling is the top-most concern among users of VPNs. But in TR top-most priority is to exclude potentially malicious networks from participating in data routing so as to guarantee high security by not allowing any form of access to the data.

**Discursion**

The privacy afforded by VPNs was only that the communications provider assured the clients that no one else would use the same circuit. This allowed clients to have their own IP addressing and their own security policies. A leased circuit ran through one or more communications switches, any of which could be compromised by someone wanting to observe the network traffic.

The VPN customer trusted the VPN provider to maintain the integrity of the circuits and to use the best available business practices to avoid snooping of the network traffic. But a VPN-based networks implementation can be attacked in many ways. Some of the possible types of VPN attacks are attacks against VPN protocols, cryptanalysis attacks and denial-of -Service attacks. And as VPN traffic is often invisible to IDS monitoring when the IDS probe is outside the VPN server, as is often the case, then the IDS cannot see the traffic within the VPN tunnel because it is encrypted. And as such if an intruder gains access to the VPN, the intruder can attack the internal systems without being picked up by the IDS.

Furthermore, in VPN, PPTP and even L2TP do not provide robust security mechanisms against rogue ISPs and tunnel routers. Though the advanced and comprehensive encryption capability provided by IPsec does offer a high degree of protection against these rogue elements yet if node replication which is the process of incrementally copying, or replicating, data that belongs to a client node is in effect, data is replicated and possibly decrypted. This kind replication can be avoided using TR method that allows excluding untrusted networks and networks situated in an unwanted locations.

VPN though provides confidentiality such that even if the network traffic is sniffed at the packet level, an intruder would only see encrypted data. But the fact that the encrypted data is available to the intruder is also a big treat as the intruder could one way or the other gets encrypted data decrypted. TR on the other hand does not allow an intruder to have access to the data at all by bypassing the intruder's network, and in the case of when the intruder trying to penetrate already trusted network, TR detects changes in the RT and thus stops data transfer with immediate effect.

Also in VPN, the Internet connectivity and the subsequent tunneling are dependent upon ISP i.e. it is usually impossible for a user to know the paths used by VPNs, or even to validate that a trusted VPN is in place; they must trust their provider completely. And as the data tunnel is rendered across multiple routers, rogue tunnel can be established and an accompanying rogue gateway. Being part of the tunnel, these intermediate routers can examine/copy/modify the data should it be malicious even if the packets are encrypted. In such a case, the sensitive user data may be tunneled to the imposter's gateway. Thus data, which is encrypted between VPN client and the VPN endpoint, can be seen while traversing public networks from sender to receiver.

In essence, merely using VPN does not give any guarantee of security but TR method does, as it allows for security measures as in the case of VPN but further allows for choice of trusted RD among available ones within the public networks like Internet. TR controls traffic flow against intrusion and ensures that data traverses only trusted routes from sender to receiver, and it provides for secured data transfer in a different way compared to VPN.

**Conclusion**

A security policy is a set of rules, practices, and procedures dictating how sensitive information is managed, protected, and distributed, thus if security policy by either the sender or the receiver stipulates that the data should not be available to third party/routing devices in certain locations, whether encrypted or not, TR method is preferable for secure data transfer from a sender to a receiver since when using TR method, data is not available to intruder(s), either encrypted or not. And this is of paramount importance as then the intruder has no data to examine/copy/

modify. Thus data being transferred using TR method is truly secured.

## References

1. ISO 7498-2-99, Information technology, Open Systems Interconnection, Basic Reference Model Part 2 - Information security architecture.

2. Mark Lewis , Comparing, Designing, and Deploying VPNs, Cisco Press, April 12, 2006

3. Tiamiyu A. Osuolale, On the Simulation of Trusted Routing Mechanism./ Topical Issues on Information and Telecommunications in Science and Education SPbSUT 2013 pp. 879-882.

4. Roy Hills, Common VPN Security Flaws, NTA Monitor Ltd, Jan 2005.

5. M.V. Buynevich, A.E. Magon, D.M Shiryaev, "The analysis of possibility of safe scaling of telecommunication structure ACS by forced routing of a traffic by standard means", Questions of a modern science and practice, V.I. Vernadsky University. Technical science series - vol. 2, no. 3(13), 2008, pp. 161-164 (Буйневич М.В., Магон А.Е., Ширяев Д.М. Анализ возможности безопасного масштабирования телекоммуникационной структуры АСУ путем принудительной маршрутизации трафика стандартными средствами.// Вопросы современной науки и практики. Университет им. В.И. Вернадского.- 2008, № 3(13).– Том 2. Серия «Технические науки».– С. 161–164).

6. Rick Graziani and Allan Johnson, routing protocols and Concepts, cisco 2011.